\documentclass{article}
\usepackage{url} %this package should fix any errors with URLs in refs.
\usepackage{lineno}
\usepackage{authblk}
\usepackage{graphicx}
\usepackage{fullpage}
\usepackage{todonotes}
\usepackage{setspace}

\usepackage{soul}
\usepackage{amsmath, bm}
\usepackage{gensymb}

\nolinenumbers

\title{The Ceiling Height of Wildland Fire Plumes in Sheared Boundary Layer Flow}

\author[1]{Jie Sun\thanks{corresponding author: Jie Sun, email: jsun2@fsu.edu}}
\author[2,3]{Kevin Speer}
\author[2,3]{Bryan Quaife}
\author[1,2]{Ming Cai}
\author[4]{David Schvartzman}

\affil[1]{\footnotesize Department of Earth, Ocean, and Atmospheric
Science, Florida State University, Tallahassee, FL, USA}
\affil[2]{\footnotesize Geophysical Fluid Dynamics Institute, Florida State University, Tallahassee, FL, USA}
\affil[3]{\footnotesize Department of Scientific Computing, Florida State University, Tallahassee, FL, USA}
\affil[4]{\footnotesize School of Meteorology and Advanced Radar Research Center, The University of Oklahoma, Norman, OK, USA}

\begin{document}
\maketitle

\begin{abstract}

Radar observations from a prescribed fire experiment reveal 
a large-scale, billow-like vorticity pair associated with 
the plume head at the onset of plume bending. 
The bending confines the ceiling height of the plume, delaying its smoke dispersion and increasing 
fire spotting risks.
This study aims to investigate the onset of plume bending in a sheared
crossflow and stratified atmospheric conditions, providing
insights into smoke dispersion and fire behavior. 
Large Eddy Simulations (LES) using the Cloud Model 1 (CM1) 
are conducted to simulate the observed development of plume 
structure and its associated dynamical fields, 
with particular focus on the plume head and its evolution from
initial plume development under different fire intensities 
and atmospheric boundary layer (ABL) conditions.
A scaling analysis of plume ceiling height is proposed based on a modified Byram’s
convective number that accounts for sheared crossflow.
The proposed scaling agrees well with the LES results,
highlighting the critical roles of shear and stratification 
in controlling plume dynamics.

\end{abstract}

\section*{Plain Language Summary}
Wildland fire burns large areas every year and releases enormous
amounts of smoke and particulates, impacting public health.
In conditions with background atmospheric wind, typically
characterized by vertical shear within the planetary boundary
layer, smoke plumes from wildland fires sharply bend over just
downstream of the fire. This effect brings the plume head
and base closer to the ground leading to stronger interactions
with surface winds and higher surface smoke concentrations.
The dynamical interaction between the fire and atmosphere
produces large eddies and ember fallout, facilitating fire
spotting. Numerical simulations and simplified models are
used to account for these effects to predict the ceiling height and
length scales of the plume.

%%%%%%%%%%%%%%%%%%%%%%%%%%%%%%%%%%%%%%%%%%%%%%%
%
%  BODY TEXT
%
%%%%%%%%%%%%%%%%%%%%%%%%%%%%%%%%%%%%%%%%%%%%%%%

%%% Suggested section heads:
\section{Introduction}
Understanding wildland fire and plume behavior is essential for 
effective firefighting and safe evacuation from areas threatened 
by advancing smoke and flames. 
Gusts and turbulence can loft embers of various shapes and sizes 
through the buoyant plume 
(\cite{koo2010firebrands, thurston2017contribution}), and these 
often produce spot fires in unburned areas. 
This spotting phenomenon, often exacerbated by strong turbulent 
winds and dry conditions, can lead to rapid fire spread over 
vast distances, making containment efforts challenging. 
The role of the background boundary layer dynamics in impacting 
the development of plume has been evidenced by the study of coupled 
fire-atmosphere studies (e.g., \cite{koo2012modelling, coen2015high}). 
Thurston et al. (\cite{thurston2016simulating}) revealed boundary 
layer rolls in the lofting process and gravity wave interactions influencing surface winds and fire front spread.

In addition to spotting, the near-ground smoke concentration resulting 
from fire plumes is a significant societal concern due to the health 
risks it poses to communities (\cite{delfino2008relationship,
wegesser2009california, holstius2012birth, johnston2012estimated}). 
Near-surface smoke poses significant societal concerns, 
including health risks to communities, persistent temperature 
inversions (\cite{robock1988enhancement, robock1991surface}), 
unpredictable smoke transport patterns (\cite{lareau2015cold}), 
and reduced visibility for travelers (\cite{ashley2015driving}). 
The dispersion of smoke is profoundly influenced by the interaction 
of the wildland fire plume with the atmospheric boundary layer 
(ABL) (\cite{sun2009importance, coen2018generation, kiefer2016study}). 
In prescribed fires, usually carried out during the day, 
the mixing of smoke across the ABL may provide sufficient smoke 
dilution to mitigate hazards. 
Atmospheric stability is a fundamental aspect of this problem: 
how to keep smoke away from people (\cite{waldrop2018introduction}).
Prior studies have discussed how daytime atmospheric mixing can 
help disperse smoke and reduce its near-surface accumulation during prescribed burns 
(e.g., \cite{potter2012atmospheric_a, potter2012atmospheric_b}). 
These studies highlight the importance of considering atmospheric 
stability and mixing depth when planning prescribed fire operations.
These prescriptions are normally based on fire weather 
parameters such as the mixed layer height, dispersion index, 
and transport wind (\cite{lavdas1986atmospheric, wade2007managing,
peterson2018nwcg}). 
Practitioners have developed rules and thresholds for these 
parameters to guide burn plans and provide adequate conditions 
to successfully carry out burns in a timely manner. 
Too much or too little convective activity and transport in 
the boundary layer can shut down operations. 
Stratification and stability in the atmosphere are recognized 
both for dispersion and for the control over the ultimate 
state of large fire plumes. 
Although not usually considered directly by practitioners, 
the vertical temperature profile plays a significant role 
in fire behavior, influencing the development of thermal plumes 
and the direction of fire spread. 
While the top of the ABL may put a lid on further vertical 
development, a small fraction, about 10\%, of plumes extend 
beyond the boundary layer (\cite{kahn2008wildfire, val2010smoke}). 

To further explore how the ABL conditions impact the behavior 
of smoke plumes generated by a wildland fire, in this study, 
we analyze the dynamical structure and evolution of a bending 
smoke plume  observed with a polarimetric Doppler radar during 
a prescribed fire experiment.  
The bending is attributed to crossflow effects in the ABL. 
With the large eddy simulation (LES) in an atmospheric 
cloud-resolving model, the observed dynamical characteristics 
of the bending smoke plume are well simulated using the simplified
ABL conditions according to the observation. 
The ABL conditions include sheared crossflow and stratification.

To extend the classic scaling of plume ceiling height, which was originally 
developed without considering crossflow (e.g., \cite{turner1973buoyancy}), 
a series of numerical experiments are conducted with varying
ABL conditions and fire intensities. 
A new diagnostic approach, based on temporally averaged eddy 
kinetic energy, is introduced to define the ceiling height 
at which plume bending occurs in the presence of crossflow.
Building on the classic theory, this study proposes an advanced 
scaling for the plume ceiling height under linearly sheared 
and stratified conditions.

%
% The main text should start with an introduction. Except for short
% manuscripts (such as comments and replies), the text should be divided
% into sections, each with its own heading.

% Headings should be sentence fragments and do not begin with a
% lowercase letter or number. Examples of good headings are:

\section{Data and Methods}
% Here is text on Materials and Methods.
%
% \subsection{A descriptive heading about methods}
% More about Methods.
\subsection{Radar data}

\begin{figure}[htp]
  \centering
  \includegraphics[width=0.9\textwidth]{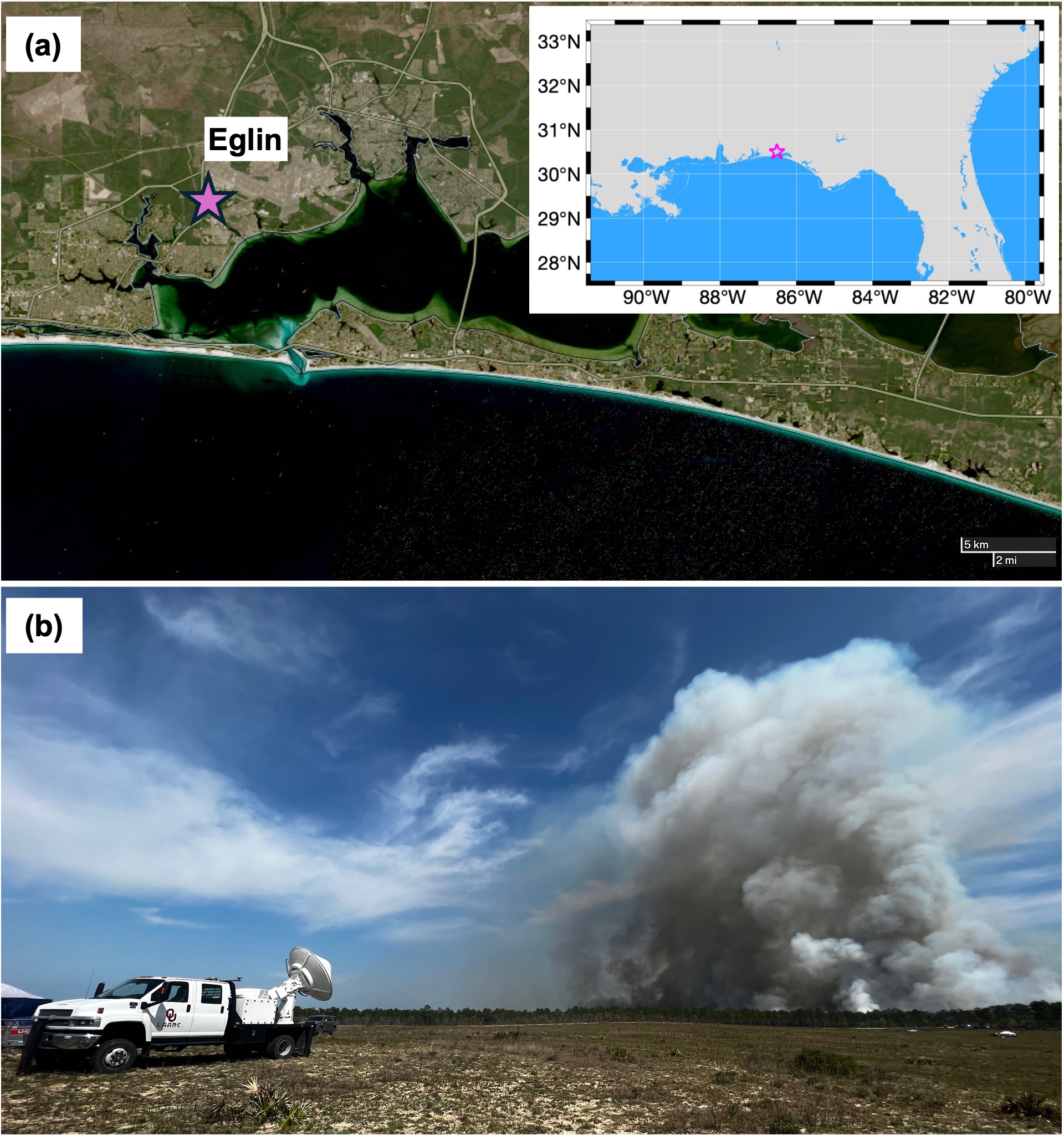}
  \caption{\label{fig1location} (a) Location of the Eglin prescribe fire on 11 March, 2023, and (b) RaXPol deployment setup. The zoomed-in map in (a) is from NASA WorldView, url{http://worldview.earthdata.nasa.gov}; last accessed June 2025. 
 }
  %  \vspace{-.5cm}
\end{figure}

The smoke plume field experiment was conducted on March 11, 2023, 
during a prescribed fire at Eglin, Florida (Fig.~\ref{fig1location}a), 
in coordination with the US Forest Service. 
Fig.~\ref{fig1location}b shows the radar deployment during 
the active plume development period. 
Observations were collected with the University of Oklahoma’s
Advanced Radar Research Center (ARRC) Rapid Scan X-band Polarimetric Radar 
(RaXPol) radar (\cite{pazmany2013mobile}). 
RaXPol is a mobile, high-resolution Doppler radar system 
designed for advanced atmospheric research.
Mounted on a truck platform, RaXPol features a dual-polarized 
parabolic dish antenna with a 2.4 meter diameter, 
capable of high-speed scanning at up to 180$^\circ$ per second. 
Operating at the X-band (3~cm wavelength), the radar provides
dual-polarization measurements~(\cite{doviak2014doppler,
Schvartzman2024radar}), including reflectivity ($Z_h$), 
Doppler velocity ($v_r$), spectrum width ($\sigma_v$), 
differential reflectivity ($Z_{DR}$), differential phase ($\Phi_{DP}$), 
and correlation coefficient ($\rho_{hv}$), 
with a range gate spacing adjustable from 7.5
to 75 meters. 

RaXPol's advanced signal processing capabilities, 
including frequency hopping and pulse compression, 
enable rapid volume scans in approximately 20 seconds, 
capturing fine-scale atmospheric phenomena. 
This system has been instrumental in observing dynamic weather events,
such as tornadoes and convective storms, 
delivering high temporal and spatial resolution data critical 
for understanding severe weather dynamics. 
For the smoke observations presented here, 
RaXPol was configured to produce range sampling of 30 m and 
with a pulse-repetition frequency of 2,000 Hz. 
Data were primarily collected in range-height indicator 
(RHI) mode, scanning through the core of the fire plume 
for approximately one hour.

%And the topography data showed the upper-right corner is downloaded from https://ngmdb.usgs.gov/topoview/.

\subsection{Atmospheric sounding data}

Sounding profiles for crossflow speed and potential temperature were 
obtained from the nearest station to the prescribed fire site 
(72230 BMX). The atmospheric sounding data were sourced from the 
University of Wyoming’s Upper Air Sounding Archive 
(\url{https://weather.uwyo.edu/upperair/sounding.html}). 
The 00Z measurement on 
March 12, 2023, the closest in time to the prescribed fire experiment 
(around 19Z on March 11, 2023), was used. 
Near-surface (below 200 m) wind profiles were obtained 
with a Remtech PA-XS SODAR, which can be found at: 
\url{https://github.com/Jie-Sun-TC-FIRE/RADAR_SODAR_Data/blob/main/SODAR_20230311.dat}.

For validation purpose, we also analyzed the ERA5 reanalysis data 
(\cite{hersbach2023era5}) on 19Z, March 11, 2023. 
Specifically, the vertical profiles of temperature 
(used to calculate potential temperature), horizontal winds 
at each pressure level for grid point (30.50°N, 86.50°W) are used, 
which is the most neighboring grid point around the burning field 
(about 30.48°N, 86.51°W). 
From the comparison illustrated in Fig. 3a, one can tell that 
the sounding profiles agree well with the reanalysis data. 
Since the sounding profiles have finer vertical resolution, 
we here use the sounding profiles as reference. 
It should be noted that, although the atmospheric sounding 
measurement was taken approximately five hours after the fire 
ignition from a site that is about 200 miles away, 
it was the closest available in both time and location. 
The data serve to indicate a typical well-mixed boundary 
layer for afternoon conditions, providing a reasonable 
estimate of the atmospheric state at the fire site, 
though not capturing the exact details of the vertical 
temperature and wind speed profiles. 
We will discuss the role of various atmospheric conditions 
in the following numerical studies.

\subsection{Numerical simulations}

The cloud-resolving model, Cloud Model 1 (CM1; \cite{bryan2002CM1}), 
is used to simulate plume evolution and compare the simulated 
results with observations to assess the model's capability in 
representing fire-generated plumes.  
A series of numerical simulations are conducted to investigate 
the role of different fire–atmosphere factors in driving plume behavior. 
The horizontal domain size in all experiments is 4 km × 4 km 
(-2000 m to 2000 m in both x and y direction) with a horizontal 
resolution of 20 m. 
The vertical domain is 0 to 12.5 km, with a stretched vertical 
resolution of 2.5–250 m (\cite{wilhelmson1982simulation}). 
The heat source of the fire is represented as a constant value 
of sensible heat flux at the surface of the domain (z = 0) 
throughout the entire simulation period. 
The domain of added heat source ranges from -520 m to -480 m 
and -1000 m to 1000 m (40 m × 2000 m) in the x and y directions, 
respectively. 
The Smagorinsky turbulent kinetic energy subgrid-scale turbulence 
closure scheme (\cite{stevens1999large}) is used in all simulations 
and no Coriolis effects are considered. 
For simplicity, completely dry simulations are applied to all 
numerical experiments, and the radiation module is turned off. 
All numerical experiments are discretized with the CM1 default 
second-order Runge-Kutta time differencing method and fifth-order 
discretizations of spatial derivatives. 
A zero-flux condition is used for the lower and upper boundaries, 
while open-radiative conditions are used for the lateral boundaries.

The role of background atmospheric conditions on plume development 
are the major concern in this study. Particularly, the effects of the vertical crossflow shear and atmospheric stratification within the boundary layer are quantitatively explored. The background atmospheric conditions are obtained from the aforementioned sounding profile. 
Specifically, a linearly increasing crossflow is applied from 
the surface to 2 km, with a constant flow above 2 km, because both
observed and simulated plumes are far below 2 km. We verified  that the background wind speed above 2 km does not affect the results of this study (not shown). This setup mimics the idealized condition of relatively uniform wind speed above the boundary layer in the absence of a  significant weather system.  Moreover, our analysis focuses on plume behavior within or near the boundary layer. 
The rate of increase varies across different experiments. A well-mixed boundary layer condition (close to neutral stratification) is applied for the lowest 1 km, while the atmosphere above 1 km is quite stable.

\section{Results} 
\subsection{Radar observed plume development}

\begin{figure}[htp]
  \centering
  \includegraphics[width=0.99\textwidth]{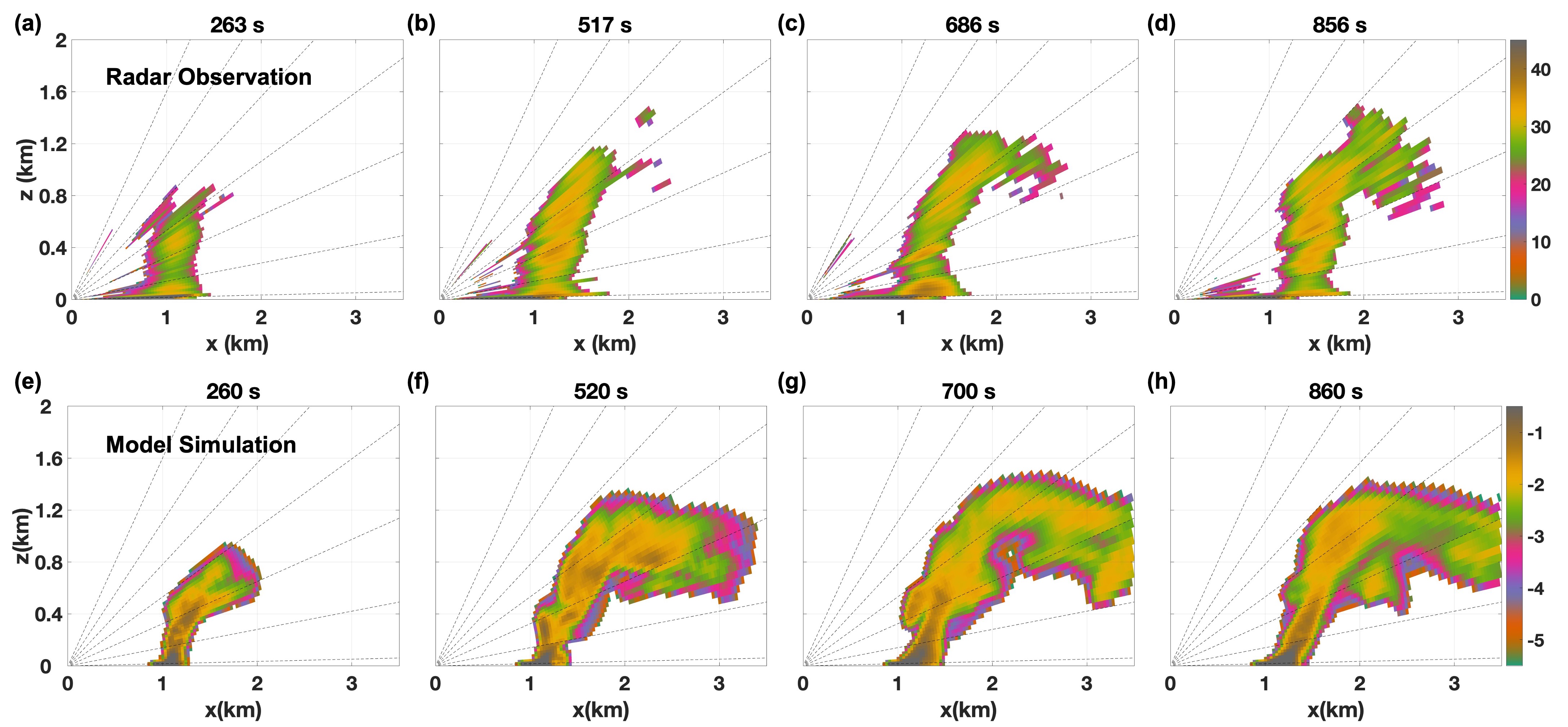}
  \caption{\label{fig2plume} The time evolution of a)–d) radar-observed 
  reflectivity (dBZ, shading) of the smoke plume, and e)–h) the simulated 
  passive tracer concentration with the CM1 model. 
  The colorbar for the passive tracer concentration is in a logarithmic 
  scale and normalized so that its maximum value is one. 
  The dashed lines illustrate the sampled radar beam lines.}
  %  \vspace{-.5cm}
\end{figure}

By vertically scanning toward the center of the plume from its 
origin for approximately 20 minutes, 
the radar observation captured the plume's development with a 
particular focus on the early development stage. 
The scan direction was closely aligned with the downstream 
background crossflow.  Among the radar’s polarization measurements, 
reflectivity ($Zh$), which typically quantifies the size and intensity 
of hydrometeors in precipitation systems, can be used to represent 
the structure of the smoke plume. 
The figures in the top panels of Fig.~\ref{fig2plume} illustrate the temporal evolution 
of the plume structure over approximately the first 15 minutes in a 
vertical cross-section (x-z plane) of radar reflectivity, 
where the z direction is vertical and x direction is aligned with the 
radar beam’s azimuth angle. 
The dashed lines in Fig.~\ref{fig2plume} illustrate the sampled radar beam lines. 
During the first several minutes (Figs.~\ref{fig2plume}a and~\ref{fig2plume}b), 
the plume showed 
a rapid vertical development represented by a quick convection.  
The plume head reached a height of 1.2 km about 8.5 minutes after the 
prescribed fire is ignited (Fig.~\ref{fig2plume}b). 
It can be indicated from Figs.~\ref{fig2plume}a and~\ref{fig2plume}b 
that the source of fire is 
located approximately 1 km from the radar (at about x = 1 km). 
Figs.~\ref{fig2plume}b–\ref{fig2plume}d display the bending structure
toward the positive x direction, 
which begins to form at the front of the observed smoke plume approximately 
8 minutes after ignition, after which the plume head moves downward. 
This bending structure, shaped much like a wave billow, 
emerges when the plume head reaches the highest level of the convection 
and begins to descend. 
The observed early development of the plume aligns well with the classic 
plume theory regarding the ``head of the plume,'' 
which exhibits hybrid features: an isolated ``thermal'' at the front of the 
plume and a steady-state plume trailing below the head 
(\cite{turner1973buoyancy}). 
The well-organized, large-scale billow structure is the key characteristic 
of this ``thermal.'' 
In Fig.~\ref{fig2plume}d, a more pronounced billow shape is observed, 
with vertical and horizontal scales comparable to the plume height. 
About 15 minutes after ignition, the plume turned to a steady state, 
no obvious large eddy vortex structure can be clearly seen at the head of the plume. 
The ``plume head'' structure drives the smoke inside the plume closer to 
the surface, even at the considerable horizontal distance of 1.5 km 
from the fire source. 
Due to the vertical circulation, the billow structure potentially leads to 
two dangerous consequences: 1) embers are ejected and land at a significant 
distance from the fire source, potentially causing spotting spread of the fire; 
and 2) smoke may be transported to the ground at a significant distance from 
the fire, posing health risks and hazards to traffic. 
Therefore, understanding what controls the height of the plume bending 
structure is crucial.

In studies of idealized vortex fields of fire-induced plume with 
a background crossflow (\cite{fric1994vortical, cunningham2005coherent}), 
a horizontal vortex tube along the background flow is typically observed 
at the plume head, corresponding to plume bending. 
This feature is referred to as part of the well-recognized 
``counter-rotating vortex pair''. 
This bending shape is characterized by a vortex tube parallel to the 
x direction, with rotation in the y-z plane and vorticity aligned with 
the background horizontal flow in the x direction. 
However, it is difficult to diagnose the “counter-rotating vortex pair” 
when the radar is aligned in the same direction. 
Nevertheless, radar observations indicate that the plume bending exhibits 
a corresponding billow-like structure. 
This structure is characterized by a clockwise circulation in the x-z plane. 
This billow-like vortex pair closely resembles the “head of the plume” 
structure in literature of classic plume theory 
(\cite{turner1973buoyancy, briggs1969plume}) and marks the height where 
plume bending occurs.

\begin{figure}[htp]
  \centering
  \includegraphics[width=0.99\textwidth]{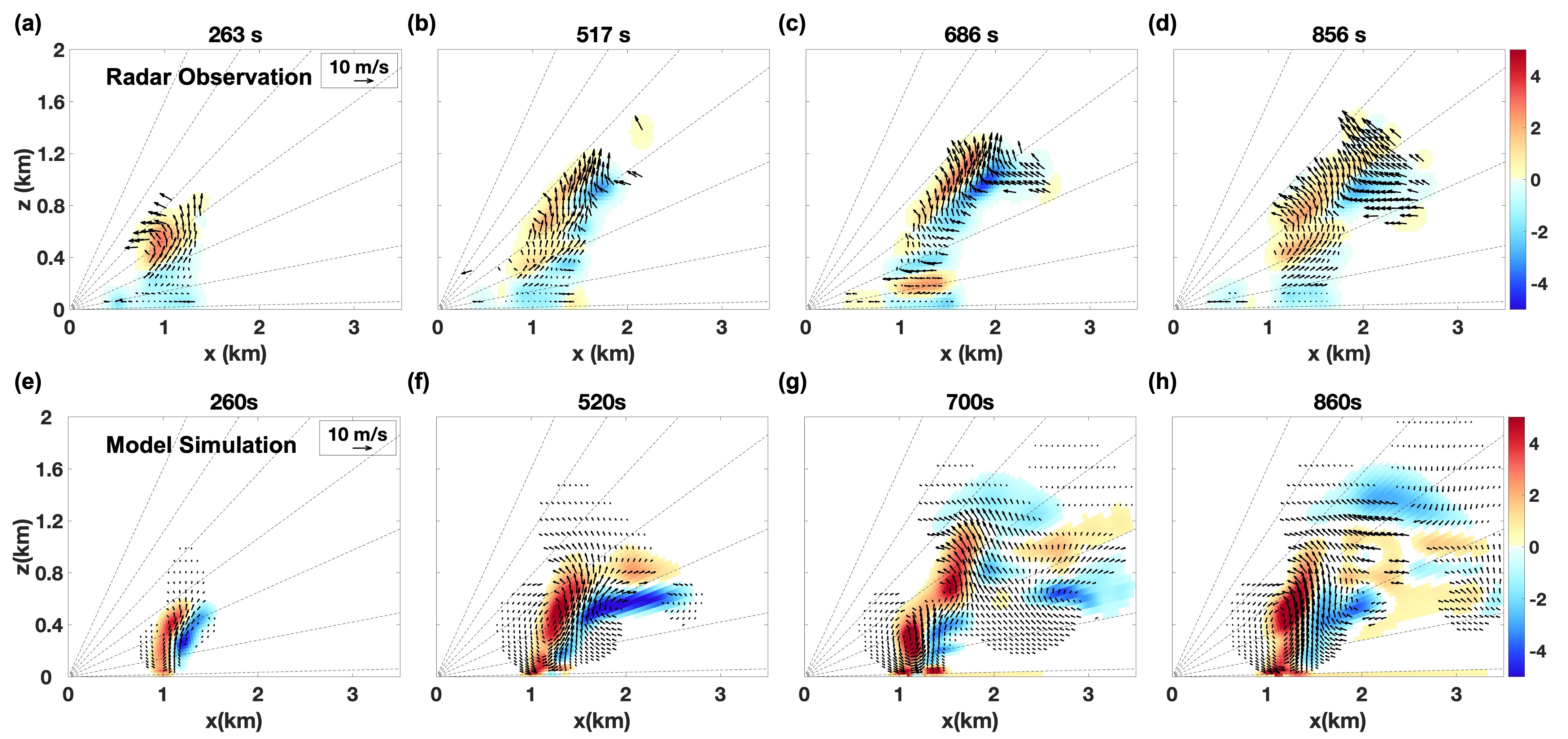}
  \caption{\label{fig3vort} The time evolution of vector winds (vectors) and
vorticity fields (shading) in $x$-$z$ plane cross section. The top row (panels a - d)
are derived from the radar-observed Doppler velocity and the bottom row (panels e - h)
are from the CM1 simulation. The background horizontal flow is removed
before calculating the vector winds and vorticity. Note that in the 
simulated velocity vector field, wind speeds below 0.5 m/s are not displayed in order to highlight the dominant structures associated with large eddies and facilitate comparison with radar observations.}
  %  \vspace{-.5cm}
\end{figure}

The development of vortices associated with the smoke plume is evident 
in radar observations (top panels of Fig.~\ref{fig3vort}). 
For clarity, we define vorticity with counterclockwise circulation in 
the $x-z$ plane as positive:

\begin{align}
    \label{eqn:vort}
  \zeta= \frac{\partial w}{\partial x} - \frac{\partial u}{\partial z},
\end{align}

where $w$ is the vertical velocity and $u$ is the velocity in the azimuthal 
direction of the radar observation ($x$ direction). 
These velocity components are calculated by decomposing the radar-observed 
Doppler velocity into the $x$ and $z$ directions. 
Specifically, if a Doppler velocity $V$ is observed along a beam line 
toward elevation angle $\varphi$, then

\begin{align}
  w = V \sin\varphi \quad \text{and} \quad
  u = V \cos\varphi.
\end{align}

It should be noted that a left-handed coordinate system is 
followed in this definition. 
However, for the purpose of displaying cross-sectional 
vorticity in the 2D $x–z$ domain, counter-clockwise circulation 
is defined as positive vorticity. 
This choice is made to align with conventional understanding.
It also should be pointed out that the radar observed velocities 
are the “Doppler velocity”, i.e., the component of the velocity 
along the radar beam direction. 
When the velocity of the plume particles are largely along the 
radar beam, the differences between the observed Doppler velocity 
and the full velocity will be small. 
The observed Doppler velocity is then decomposed into 
$w$ and $u$ components. 
Besides the small-scale  turbulence-like eddy fields in the top 
panel of Fig.~\ref{fig3vort}, a well-organized vortex structure 
is also evident throughout the plume’s development. 
This large-scale vortex field consists of a positive vorticity 
(counterclockwise) structure on the upstream side of the plume 
and a negative vorticity (clockwise) structure on the downstream side. 
This configuration aligns with the classic plume model 
(\cite{fric1994vortical, cunningham2005coherent}).

At the onset of the fire, the vortex pair primarily propagates upward 
associated with the rapid convection. 
When the top of the convective plume reaches its maximum height 
(approximately 1.4 km) about 10 minutes after ignition, 
the horizontal movement of the vortex pair becomes increasingly 
significant (Fig.~\ref{fig3vort}b and~\ref{fig3vort}c). 
This horizontal motion causes the plume to appear as if it reflects 
at the top of the highest convection level. 
In the presence of background horizontal flow, the positive and negative 
vortex pair couples to form a billow-like structure, with positive 
vorticity at the top and negative vorticity at the bottom. 
This structure recurs and is similar to the billows seen in the 
sea-breeze phenomenon but distinct from the horizontal 
“counter-rotating vortex pair” tube previously mentioned. 
The downward motion of this large-scale billow structure presents 
a significant potential of hazard. 
The clockwise circulation within the billow can project embers 
from the top of the plume to distances far beyond the plume’s 
original reach. 
Additionally, the low elevation of the billow’s base and the 
downward flow in front of the billow can bring smoke to the 
ground downwind of the fire, increasing the risk to people 
and communities.

\begin{figure}[htp]
  \centering
  \includegraphics[width=0.99\textwidth]{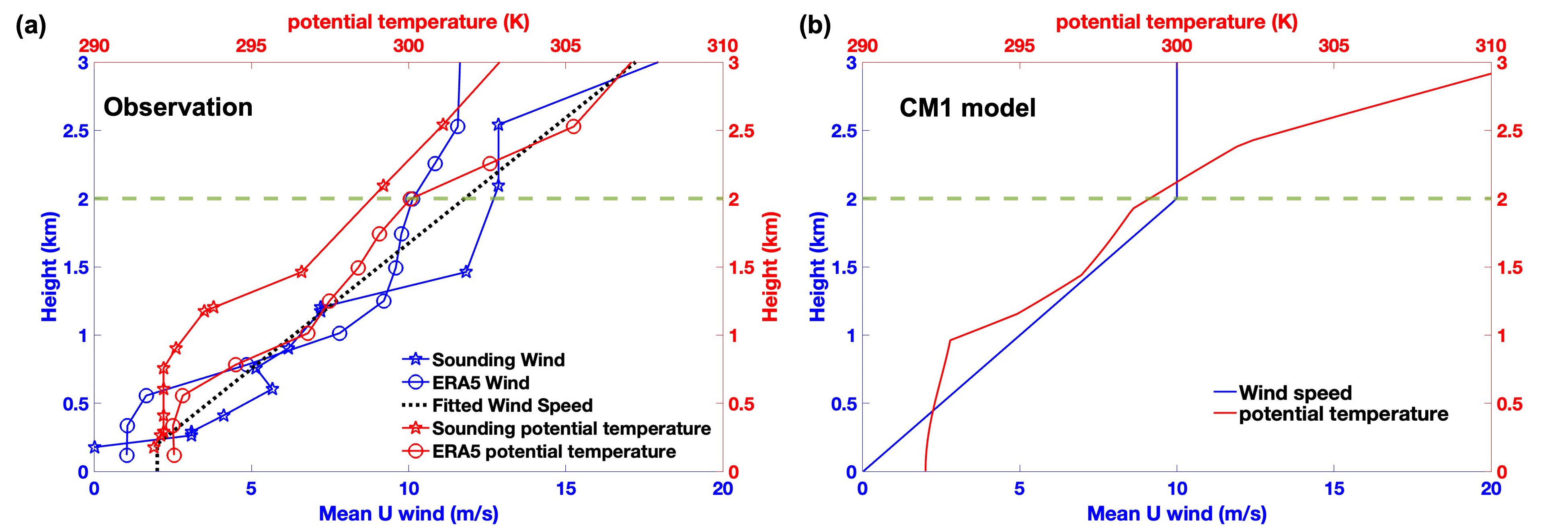}
  \caption{\label{fig4sounding} Vertical profiles of the background wind
speed (blue curves) and the potential temperature (red curves). In (a), the 
curves with stars represents sounding profiles from station 72230 BMX and 
curves with circles represents vertical profiles derived from ERA5 data 
at the location most close to the fire field.  
In (b), the curves represent the vertical profiles used as background in CM1 
simulation. The dashed black curve in (a) represents the linearly 
fitted wind speed profile within the lowest 3 km, based on the sounding. 
A constant wind speed of 2~m/s is prescribed in the lowest 200~m for 
the dashed black curve, as indicated by the sodar observations.}
  %  \vspace{-.5cm}
\end{figure}

\subsection{Modeling of the smoke plume development}

The previously described CM1 model is used to simulate the 
fire generated plume and further investigate the factors 
influencing the height of the plume bending. 
First, the simulation is designed to represent the observed 
plume and its bending structure associated with the billow-like 
large eddies under the observed environmental conditions, 
which is referred to as the “Control Run.” 
Specifically, the sounding and reanalysis data are used as 
reference for the simplified vertical profiles of the background 
potential temperature and the horizontal winds. 
Fig.~\ref{fig4sounding}a shows a well-mixed boundary layer with a 
nearly constant potential temperature of approximately 292 K in 
the lowest 1 km, which is conducive to plume convection development. 
The background horizontal wind increases linearly with height. 
For simplicity, in the numerical model, an unchanged background wind 
blowing  toward the positive $x$-direction 
is applied throughout the entire simulation, 
with wind speed increasing  linearly from 0 m/s at the surface 
to 10 m/s at 2 km (Fig.~\ref{fig4sounding}b). 
The background thermal stratification in the simulation is represented 
by a well-mixed layer below 1 km, transitioning to a linearly 
increasing potential temperature profile above 1 km. 
The background conditions used in the numerical simulation 
(Fig.~\ref{fig4sounding}b) are in close agreement 
with the observed profiles.

\begin{figure}[htp]
  \centering
  \includegraphics[width=0.99\textwidth]{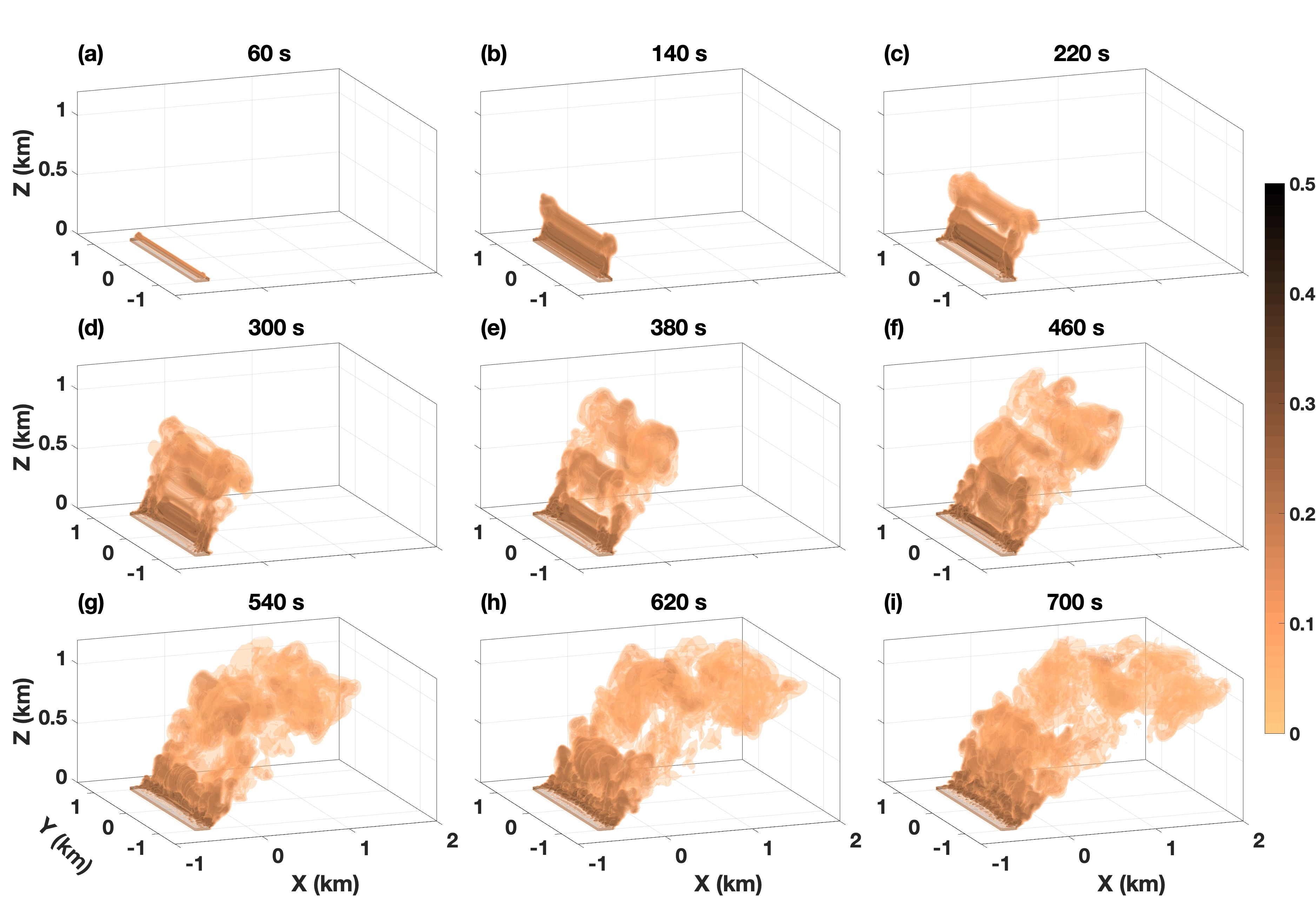}
  \caption{\label{fig5smoke} The time evolution of the CM1 simulated
  three-dimensional smoke plume represented by a passive
  tracer concentration. The color label represents the concentration of
  the passive tracer normalized by the concentration at the source.}
  %  \vspace{-.5cm}
\end{figure}

To mimic the smoke source in the control run, a spatially uniform and 
temporally constant flux of passive tracer is added just above the fire 
source region (Fig.~\ref{fig5smoke}) throughout the entire simulation. 
The added passive tracer keeps a unitless concentration (intensity) of 1 
throughout the entire simulation. 
The resulting spatiotemporal distribution of the tracer field is also 
unit-less, representing relative intensity with respect to the source. 
It is used as a proxy for smoke concentration and illustrates the shape 
of the smoke plume. 
In the first minute of the simulation (Fig.~\ref{fig5smoke}a), 
plume convection is triggered by the buoyancy force induced from the 
surface heating flux. 
As convection develops, a well-organized vortex tube structure emerges. 
From 60 s to 460 s (Fig.~\ref{fig5smoke}b–\ref{fig5smoke}f), 
the vortex tube at the front of the plume remains largely aligned with 
the y-direction, serving as a main characteristic of the plume head.
The axisymmetric tube structure suggests that the plume’s 3D spatial 
structure can be approximated in a 2D $x-z$ plane, making it easier to 
compare with radar-observed 2D plume data. 
In the cross-section profile of this tube along a constant y (i.e., $y = 0$),
a billow-like vortex pair appears in the $x-z$ plane 
(bottom panels of Fig.~\ref{fig2plume} and Fig.~\ref{fig3vort}). 
At time 620 s (Fig.~\ref{fig5smoke}h), a bending structure is evident 
at the front of the plume, 
just below the highest convective level, and begins to move downstream 
with the crossflow. 
At later times, the well-organized vortex tube structure transitions 
into small-scale turbulence, suggesting that the billow-like vortex pair 
is significant only during the first one or two large-scale plume eddies 
generated by the simulated fire. 
Real fires would naturally produce numerous start-up billows as the fire 
waxes and wanes across inhomogeneous fuels, topography, 
and with variable weather.

Next, the 2D development of the simulated plume will be compared with the 
observational counterpart. 
In the bottom panels of Fig.~\ref{fig2plume}, the vertical cross-sections 
of the model’s passive tracer concentration field are displayed, 
which mimics the smoke concentration to be compared with the radar 
reflectivity field. 
By comparing the top and bottom panels of Fig.~\ref{fig2plume}, it is 
evident that our model successfully represents the smoke plume’s 
development, including the behavior of the smoke particles 
within the plume. 
The simulation also shows that the bending structure promotes the 
movement of smoke from the top of the plume toward the ground, 
extending the distance the smoke travels away from the fire source.
The timing of the simulated plume development can also match well 
with the radar observation.

Then the spatiotemporal development of 2D vorticity field of the simulated 
plume (bottom panels of Fig.~\ref{fig3vort}) is compared with the observations. 
In each of the bottom panels, the simulations are smoothed using a spatial 2D 
Gaussian smoothing kernel with a standard deviation of 1. 
The spatial smoothing helps to highlight the large-scale features of 
the billow vortex structure in the $x-z$ plane. 
In the numerical simulation, the first billow-like vortex pair develops 
at the downstream of the prescribed surface heating (Fig.~\ref{fig3vort}f) 
because of the existence of crossflow. 
The vortex pair’s location (about 1.5 km from the heating source), 
time (around 9 minutes), and width (approximately 1 km) are well simulated 
compared with the radar observation (c.f. the top and bottom panels of 
Fig.~\ref{fig3vort}). 
The front edge of the first billow exhibits counterclockwise circulation 
aloft and upstream, and clockwise circulation below and downstream. 
The simulation also reveals finer turbulent flow structures, featuring 
multiple paired vortices with positive vorticity aloft and towards 
the back, and negative vortices below and towards the front, 
all within a single billow. 
A key feature, visible in the bottom row of Fig.~\ref{fig3vort}f, 
is the well-organized negative vorticity vortex, which is significantly 
lower than the highest convective level and much closer to 
the ground (as low as 500 m). 
The horizontal distance of the billow is slightly larger than 1 km 
from the fire location. 
The existence and location of the billow structure are consistent 
with radar observations.

About 700 s after ignition, another billow-like vortex pair forms 
at nearly the same location as the first one, with a similar width, 
downstream of the stationary heating source (Fig.~\ref{fig3vort}g). 
Meanwhile, the first billow has propagated farther downstream and 
become less organized (Fig.~\ref{fig3vort}g). 
As time progresses, vortices are continuously generated from the 
heating source, following the previously formed billows as they move 
upward and then bend downstream with the crossflow. 
During this process, the vortices gradually dissipate and transition 
into turbulence, while the main body of the bent-over plume evolves 
toward a quasi-steady state (Fig.~\ref{fig3vort}h). 
The vorticity field associated with this quasi-steady plume structure 
also aligns well with the observational study of \cite{lareau2017mean}. 
In reality, the downstream advancement of the fire can lead to the 
continuous lifting of embers into the smoke plume. 
Subsequently, the continuous formation of billow-like vortices 
at the bending point of the plume can bring embers to the ground 
at significant distances from the fire front, playing a crucial 
role in the rapid, ongoing spotting spread of fires as evidenced in 
\cite{lareau2025plume}. 
 
In summary, the simulation conducted with CM1 demonstrates 
close agreement with observations, both for the plume structure 
and the internal dynamics. 
This consistency indicates that when the background atmospheric 
conditions provided to CM1 match the observed conditions, 
the model accurately represents the dynamics that drive fire plume 
development. 
Consequently, additional simulations are designed to explore the 
mechanisms of vortex formation and quantify their propagation 
in the smoke plume under different atmospheric conditions and 
heating sources.

\begin{table}
 \caption{\label{tbl:experiments} Numerical experiments for scaling of plume height. $Q$ is the
heating rate at the surface, and ``Shear" is the vertical change of
background horizontal flow from surface to 2~km height. The bold numbers
correspond to the ``control run" that replicates the radar
observations.}
\begin{center}
 \begin{tabular}{|l|c|c|c|c|c|}
 \hline
  Exps. & $Q$ (kW m$^{-2}$) & $N$ (s$^{-1}$) & Shear (m s$^{-1}$ over 2~km)   \\\hline
   Control & 30 & 0.006 &  10   \\\hline
   Group 1 & 30 & 0.006 & 5, 10, 15, 20, 25, 30   \\\hline
   Group 2 & 30 & 0.012 & 5, 10, 15, 20, 25, 30   \\\hline
   Group 3 & 50 & 0.006 & 5, 10, 15, 20, 25, 30   \\\hline
   Group 4 & 20 & 0.006 & 5, 10, 15, 20, 25, 30   \\\hline
 \end{tabular}
\end{center}
\end{table}

\subsection{Scaling analysis of plume ceiling height in a stratified 
boundary layer with linearly sheared crossflow}

Using the background conditions in the ABL collected from observational data, 
the previous section demonstrates that the CM1 model can well simulate 
the observed plume development. 
In addition to the atmospheric stratification and vertical shear of crossflow, 
the intensity of the fire source plays a crucial role, 
as it determines the convective buoyancy and, consequently, 
the ceiling height of the plume. 
In this section, an extended scaling of the ceiling height of plume under 
various conditions of ABL will be derived.

A series of numerical experiments is conducted to investigate the influence 
of three key factors, including atmospheric stratification, vertical shear crossflow, 
and surface heating intensity, representing varying boundary layer conditions 
and fire sources, on the ceiling height of the plume. 
Four groups of experiments are conducted, and summarized in Table 1. 
Within each group, six experiments are performed with varying values of 
crossflow shear, corresponding to vertical changes in crossflow from the 
surface (at rest) to 2 km altitude, with speeds of 5 m/s, 10 m/s, . . . , 
30 m/s. 
In Group 1, the fire intensity and atmospheric stratification are set 
identical to the control run. 
In Group 2, conditions are the same as in Group 1, except that the 
stratification in the lowest 1 km is doubled. 
In Groups 3 and 4, conditions remain the same as in Group 1, 
except that the surface heating flux is set to 50 and 20 $kW m^{-2}$, 
respectively, to represent variations in fire source strength.

\begin{figure}[htp]
  \centering
  \includegraphics[width=0.99\textwidth]{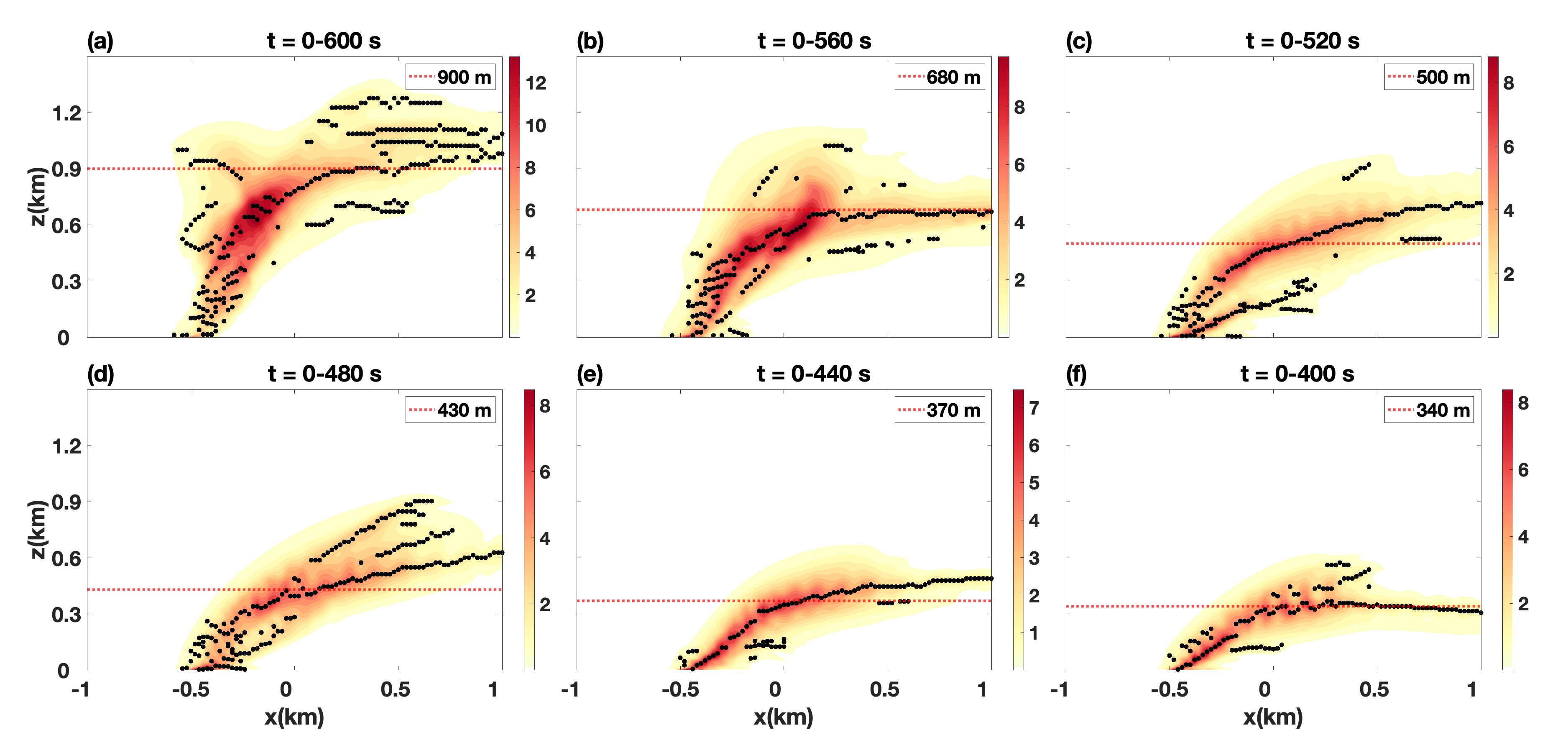}
  \caption{\label{fig6ztop} Time-averaged eddy kinetic energy (EKE, shadings)
  for the simulations from Group 1. 
  Each panel represents the time-averaged EKE in the $x$-$z$ domain for the 
  shear velocities 5~m s$^{-1}$, 10~m s$^{-1}$, \ldots 30 m s$^{-1}$ over 
  the vertical 2 km domain. 
  The black dots represent the location of local maxima in the vertical domain 
  at each horizontal grid.
  The horizontal dashed red lines indicate the height of plume bending with 
  its value listed in the box at the upper right corner. 
  Panel (b) is for the previously mentioned control run. 
  Note that the periods for the averaged EKE field, which are listed in the 
  title of each panel, are different in each experiment. 
  Larger shears have smaller time periods and smaller shears have larger 
  time periods. The black dots represent the local maximum EKE values in 
  the vertical profile at each horizontal ($x$) location.}
  %  \vspace{-.5cm}
\end{figure}

To objectively define the height of the plume bending, i.e., the ceiling height, 
the time-averaged eddy kinetic energy (EKE) is examined, 
as it effectively tracks the location of the vortex pair, 
which exhibits the greatest perturbed wave energy compared to the surrounding 
environment. 
EKE is defined as:
\begin{align}
  EKE = \frac{1}{2}(U^{'2} + W^{'2}),
\end{align}
where $U^{'2}$ and $W^{'2}$ are the perturbed horizontal and vertical speed, respectively.
The local maxima of EKE will indicate the location where the plume bending occurs. 
Fig.~\ref{fig6ztop} shows time-averaged EKE and the corresponding local maxima (black dots) 
for experiments under various vertical shear crossflows (Group 1), 
in which the same atmospheric stratification and surface heating rate as the control run
are used. 
Note that the averaging periods differ among these experiments, 
with higher-shear experiments using shorter averaging periods 
because plume gets bent earlier under higher wind shear.
In this study, the time span used to average EKE starts at 10 minutes for the case 
with 5~m/s shear, and is reduced by 40 seconds for each subsequent experiment in Group 1.
It should be noticed that results are not sensitive to the time span of averaging.
The time-averaged EKE for the ``control run" is displayed in Fig.~\ref{fig6ztop}b, 
where a two-branch structure of the plume is clearly discernible 
beyond the shared vertical convective plume at about $X = 200$ m and $Z = 700$ m.
The higher branch is roughly aligned in the vertical direction but remains short 
due to slower development, while the lower branch is much longer and more horizontally oriented. 
The intersection of these two branches corresponds to the bending point of the plume. 
The higher branch reflects the slower diffusion of the plume 
(Fig.~\ref{fig3vort}d and Fig.~\ref{fig3vort}g), while
the lower branch represents the bending structure led by the head of
plume and the corresponding initial large eddy vortex pair as indicated 
by Fig.~\ref{fig3vort}b and Fig.~\ref{fig3vort}e.
The height of plume bending can be 
determined by connecting the relatively horizontal local maxima points,
which is approximately 700 m for the Control Run.
Therefore, the ceiling height of plume ($Z{top}$) is identified as 680 m 
(red dashed horizontal line in Fig.~\ref{fig6ztop}b).
As a general trend, under the same heating source and atmospheric 
stratification, the plume ceiling height decreases as the vertical shear 
of crossflow increases. 
From Fig.~\ref{fig6ztop}, one can see that the plume ceiling height varies 
between 340 m and 900 m for the Group 1 experiments.

\begin{figure}[htp]
  \centering
  \includegraphics[width=0.99\textwidth]{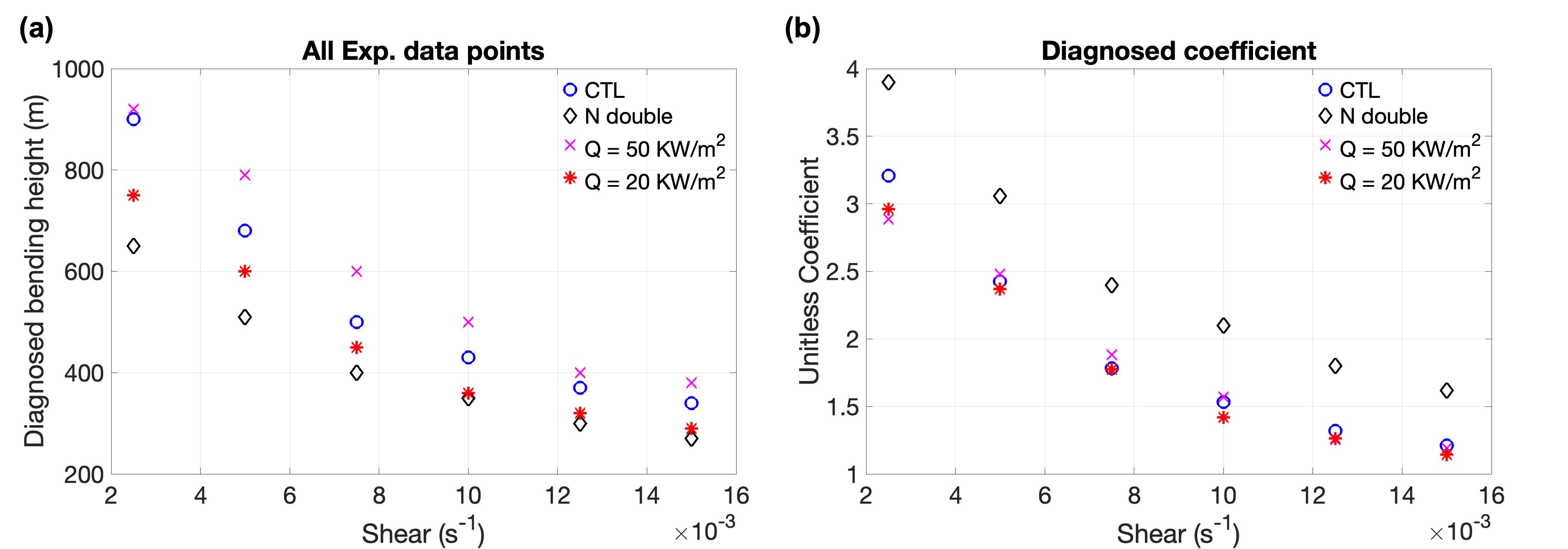}
  \caption{\label{fig7exp} The height of the plume bending for various
vertical shear crossflow, atmospheric stability, and fire source
intensity. (a) The diagnosed the plume ceiling height (ordinate) and
the vertical shear of crossflow (abscissa) for all the 24 experiments
listed in Table~\ref{tbl:experiments}. (b) Diagnosed plume ceiling height
coefficient $A_1$ (ordinate) and the vertical shear of crossflow
(abscissa). The different markers represent different experiment groups
listed in Table~\ref{tbl:experiments}. The blue circles are for Group 1 (CTL);
the black diamonds for Group 2 (N double); the magenta crosses for
Group 3 ($Q = 50~KW~m^2$); and the red asterisks for Group 4 ($Q = 20~KW~m^2$).}
  %  \vspace{-.5cm}
\end{figure}

Using the same way to determine plume ceiling heights in Fig.~\ref{fig6ztop}, 
such heights for all experiments listed in Table 1 are obtained. 
The results are shown in Fig.~\ref{fig7exp}a, with each marker 
representing an individual experiment. 
As a result, the overall relationship between the diagnosed ceiling height and the vertical shear of crossflow is evident in 
Fig.~\ref{fig7exp}a. 
In general, a stronger fire intensity, weaker crossflow shear, 
and more unstable stratification in the boundary layer result 
in a greater plume ceiling height. 
However, the relationship between $Z_{top}$ and these three factors 
is complex and nonlinear.

Classic point-source plume theory 
(e.g., \cite{turner1973buoyancy, briggs1969plume}) suggests 
that in the absence of crossflow, the ceiling height is proportional 
to the 1/4 power of the buoyancy flux ($F$ , see Appendix A for details) 
and -3/4 power of the Brunt-V\"{a}is\"{a}l\"{a}
frequency ($N$), as expressed by 
\begin{align}
  Z_{top} = 5.0 \:\left(\frac{F}{N^3}\right)^{1/4}.
\end{align}
This relationship has been validated against field experiments. 
Under the influence of vertically uniform crossflow, 
the modified scaling provides a rough dimensional estimate, i.e., 
\begin{align}
  Z_{top} \propto \left(\frac{F}{U_0 N^2}\right)^{1/3},
\end{align}
where $U_0$  is the speed of a constant crossflow. 
To date, scaling plumes' ceiling height has not been 
achieved through observational or numerical experiments. 
The new scaling of cieling height is derived based 
on numerical experiment data illustrated in Fig.~\ref{fig7exp}a. 
First, the plume ceiling height is decomposed into two components: 
the traditional scaling of plume height in the absence of crossflow 
and a modification term accounting for the influence of 
vertical shear crossflow, which is expressed mathematically as
\begin{align}
\label{eqn:ztop0}
  Z_{top} = A_1 \left(\frac{F}{N^3}\right)^{1/4}.
\end{align}
In Eq.~\eqref{eqn:ztop0}, $A_1$ is a nondimensional coefficient.
In the classic plume theory in absence of crossflow, 
the typical value of $A_1$ is approximately 5 (\cite{turner1973buoyancy}). 
When crossflow is considered, $A_1$ is not a constant anymore, 
but a function with respect to the crossflow.
Next, $A_1$ is scaled using the three predefined background 
conditions through the series of numerical experiments. 
Before proceeding, the ``diagnosed" or ``target" value of $A_1$ is 
calculated through dividing the plume height ($Z_{top}$) 
with $(F/N^3)^{1/4}$. 
The results of such $A_1$ values versus the shear of crossflow 
for all the experiments in Table~\ref{tbl:experiments} are 
illustrated in Fig.~\ref{fig7exp}b. 
The diagnosed $A_1$ ranges from approximately 1 to 4 across various
scenarios of fire intensity, stratification, and crossflow shear, which
aligns reasonably well with the expected value of 5 in the absence of
crossflow. 
The varying rates of change in $A_1$ with crossflow shear
(abscissa of Fig.~\ref{fig7exp}b) across the different experimental groups
suggest that $A_1$ depends on all three factors: crossflow shear, fire
intensity, and atmospheric stratification. 
To capture these dependencies, a modified Byram's convective number,
is introduced to incorporate the effects of all the three factors.

The original Byram's convective number 
(e.g., \cite{byram1959forest, morvan2018wildland, 
zhang2020study, morvan2024properly}) is defined as:
\begin{align}
\label{eqn:nc0}
  N_{c0} = \left(\frac{W_c}{U_c}\right)^3 = \frac{2gI}{\rho c_p \theta_0 (U_{10}-r)^3},
\end{align}
where $\frac{W_c}{U_c}$ symbolically represents the ratio of the vertical velocity 
induced by buoyancy to the background horizontal flow speed;$I = QX_0$ 
is fire intensity for a line source, with $X_0$ being the width of the fire-line;
$U_{10}$ background horizontal flow speed at 10 m height;
and $r$ is the speed of fire spread.

In this study, the Byram's convective number is modified to:
\begin{align}
\label{eqn:nc}
  N_c = \frac{W_c}{U_c} = \frac{\left(\frac{2gQX_0}{\rho_0 c_p T_0}\right)^{1/3}}{Z_c \Lambda}  = \frac{\left(\frac{2gQX_0}{\rho_0 c_p T_0}\right)^{1/3}}{\left(\frac{F}{N^3}\right)^{1/4} \Lambda}.
\end{align}

Similar to the original Byram's convective number, the modified Byram's
convective number also aims to describe the ratio between the buoyancy force
from the fire heating and the inertial force induced by the horizontal
crossflow. 
$W_c$ follows the same scaling as the original definition of $N_c$. 
However, $U_c = Z_c\Lambda$, where $\Lambda$ represents the vertical shear
rate of background flow;
thus, $U_c$ varies with the convection height, $Z_c$,
rather than being fixed at 10 m, 
as it is intended to represent the crossflow 
speed at the plume's ceiling height.
The scaling of $Z_c$ follows the classic scaling of plume height, $(F/N^3)^{1/4}$.
Thus, the new scaling can be expressed as a best-fit relationship between the 
diagnosed dimensionless variable $A_1$ (ordinate of Fig.~\ref{fig7exp}b) and 
``$N_c$": $A_1 = f(N_c)$. 
Using the MATLAB least-squares fitting function ``lsqcurvefit'', 
the best-fitting for the points illustrated in Fig.~\ref{fig7exp}b can 
be obtained as $ A_1 = 1.29 \: N_c^{1/2}$. 
Therefore, the new scaling for the plume ceiling height is:
\begin{align}
\label{eqn:ztop}
  Z_{top} = 1.29 \: N_c^{1/2}\left(\frac{F}{N^3}\right)^{1/4}   .
\end{align}

\begin{figure}[htp]
  \centering
  \includegraphics[width=0.99\textwidth]{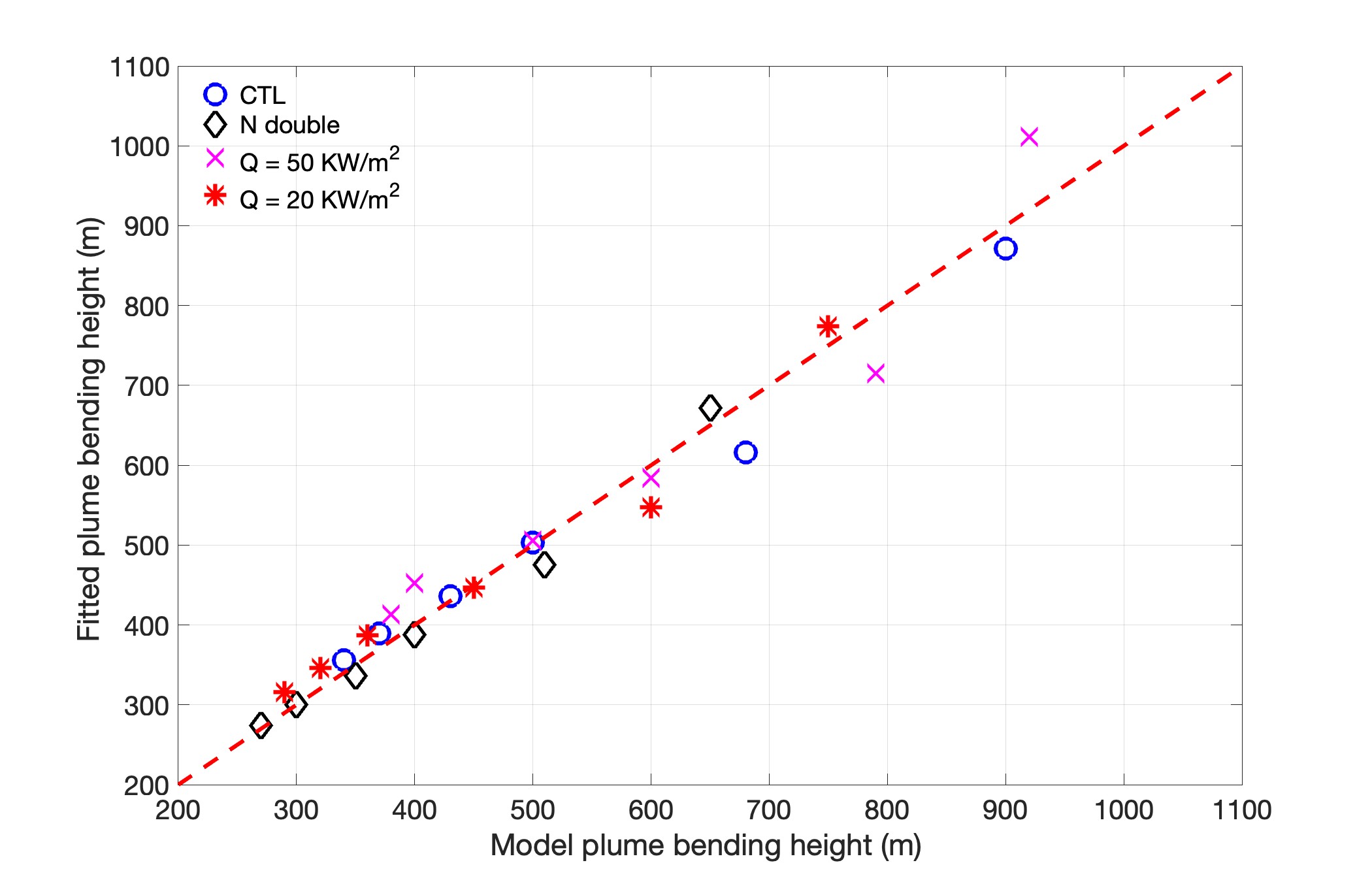}
  \caption{\label{fig8fit} The scaling results of the plume heights using Eq.~\eqref{eqn:ztop} 
  (ordinate) versus the model simulations (abscissa) for the 24 experiments listed in 
  Table~\ref{tbl:experiments}. The different markers represent different experiment groups listed in Table~\ref{tbl:experiments}. 
  The blue circles are for Group 1 (CTL),
the black diamonds for Group 2 (N double); the magenta crosses for
Group 3 ($Q = 50~KW~m^2$); and the red asterisks for Group 4 ($Q = 20~KW~m^2$).
The dashed red line is the identity line corresponding to a perfect fit.}
  %  \vspace{-.5cm}
\end{figure}

Figure~\ref{fig8fit} shows the scaling analysis results for $Z_{top}$ 
using Eq.~\eqref{eqn:ztop} versus their model simulation counterpart. 
The good agreement between the model simulated plume ceiling heights and those 
calculated from Eq.~\eqref{eqn:ztop} confirms that the new scaling 
provides a good predictive estimate for the plume ceiling heights 
under varying conditions of heating intensity, 
atmospheric stratification, and vertical shear of crossflow. 
It can be concluded that Eq.~\eqref{eqn:ztop} serves as a valid 
extension of the classic plume scaling and as an empirical prediction 
for the height of plume bending
under different atmospheric conditions and fire intensities.

%%%%%%%%%%%%%%%%%%%%%%%%%%%%%%%%%%%%%%%%%%%%%%%%%%%%%%%%%%%%%%%%%%%%%%%%
\section{Summary and conclusion}

In this study we analyze the evolution of a smoke plume generated in a
prescribed fire experiment, observed by X-band Doppler radar.  
The presence of background crossflow induces a bend in the plume, 
which limits its vertical rise and delays its
dispersion into higher atmospheric levels. 
Additionally, we identify a large-scale, well-organized, billow-like 
vortex pair co-located with the bend. 
Both the vertical and horizontal scales of this structure are closely 
related to the plume height itself. 
The billow-like vorticity field may promote the rapid spread of embers 
and
fire spotting, as well as influence near-ground smoke concentration.
Using a cloud-resolving model (CM1) with its large eddy simulation, we
well simulate the evolution of the observed plume from the
prescribed fire, particularly the large-scale, well-organized vortex
pair associated with the plume bending structure.

An approach based on the time-averaged EKE field is introduced to 
estimate the ceiling height of plume in a series of numerical experiments 
conducted under varying fire heat intensities and ABL conditions. 
Building upon the classic scaling for the plume height induced by a 
point-source fire in the absence of crossflow, 
we extend the scaling analysis to the height of plume bending generated 
by a line-source fire across these experiments.
Specifically, the proposed scaling consists of two components: 
the classic point-source plume height scale from traditional plume theory, 
and a newly introduced coefficient that captures the influence of crossflow, 
expressed as a function of the modified Byram’s convective number.

The scaling results indicate that fire intensity and atmospheric
boundary layer conditions play a crucial role in determining the height
and horizontal structure of plume and, consequently, the smoke
distribution within the boundary layer. More unstable boundary layers
promote vertical plume development, allowing smoke and billows to reach
higher altitudes. Stronger vertical sheared crossflow enhances plume
bending toward the ground and facilitates near-field fire spotting,
while smaller vertical shear allows greater vertical plume development.

%%%%%%%%%%%%%%%%%%%%%%%%%%%%%%%%%%%%%%%%%%%%%%%
%% Optional Appendices go here
%
% The \appendix command resets counters and redefines section heads
%
% After typing \appendix
%
\appendix
\section{Scaling for the buoyancy flux}

Following~(\cite{turner1973buoyancy}), the buoyancy flux is defined as 
\begin{align}
  F = \frac{Q S_0 g}{\rho_0 c_p T_0},
\end{align}
where $Q$ is the intensity of sensible heat flux from the surface with
units W m$^{-2}$, $S_0$ is the area of the heating at the surface with
units m$^{2}$, $g = 9.8$ m s$^{-2}$ is gravitational acceleration,
$\rho_0 =$ 1.2 kg m$^{-3}$ is the background air density, $c_p =$
$1.005 \times 10^3$ J kg$^{-1}$ K$^{-1}$ is the background air specific heat
capacity, and $T_0 =$ 292~K is the background air temperature at the
surface. Therefore, the units of buoyancy flux $F$ is m$^{4}$ s$^{-3}$.
In classic plume theory, a point source is used to derive the relation
between the height of plume and the buoyancy flux. In this study, $F$
is defined in terms of a line source of sensible heat flux at the
surface. Therefore, we need to quantify the area $S =
X_{_0}Y$, where $X_{_0} =$ 40~m is the fire width and $Y$ is the length
scale along the $Y$ direction which is to be determined. At the surface, we write the
length of heating of the rectangle area as $Y_{0} = \alpha X_{0}$. Then,
we can naturally assume that the ratio of the length along the $x$
and $y$ directions remain unchanged, which means that in the plume and at any
height, $Y = \alpha X$. Following the same derivation of Z$_{top}$
(i.e., Eqs.~6.1.2-6.1.4 in~\cite{turner1973buoyancy}) in a stratified
environment, we replace $\pi R^2$ with $\alpha XY$ in both sides of the
equation. This will cancel $\alpha$ in the equation and the heating area
$S_0$ in $F$ can be written as $S_0 = X_0^2$.

%%%%%%%%%%%%%%%%%%%%%%%%%%%%%%%%%%%%%%%%%%%%%%%
%
% DATA SECTION and ACKNOWLEDGMENTS
%
%%%%%%%%%%%%%%%%%%%%%%%%%%%%%%%%%%%%%%%%%%%%%%%

\section*{Open Research Section}
\textbf{Data Availability Statement}

The RaXPol radar data can be found at \url{https://radarhub.arrc.ou.edu/}. The atmospheric sounding data can be obtained from \url{https://weather.uwyo.edu/upperair/sounding.html}. 
The SODAR data for the lowest 200 m wind can be found at: 
\url{https://github.com/Jie-Sun-TC-FIRE/RADAR_SODAR_Data/blob/main/SODAR_20230311_pub.dat}.
The ERA5 data is available at 
Copernicus Climate Change Service Climate Data Store, 
\url{https://doi.org/10.24381/cds.bd0915c6}. 

\paragraph{acknowledgments}
This study is partially funded by the Department of Defense Strategic Environmental Research and Development Program (RC20-1298), and National Science Foundation (AGS-2427321, and AGS-2427323).

We thank Marcus Williams (US Forest Service) for facilitating the field deployment.

%\bibliographystyle{plain}
%\bibliography{refs}

\end{document}